\documentclass[12pt]{amsart}

\usepackage{amsmath,amssymb,enumerate}

\usepackage{epsfig,fancyhdr,color}

\usepackage{amssymb}
\usepackage{amsmath,amsthm}
\usepackage{bbm}
\usepackage{latexsym}
\usepackage{amscd}
\usepackage{psfrag}

\usepackage{graphicx}
\usepackage[all]{xy}

\usepackage[font=small,labelfont=bf]{caption}

\usepackage{mathtools}

\usepackage{tikz-cd}

\usepackage{stackrel}

\usepackage{mathabx,epsfig}

\usepackage[utf8]{inputenc}

\numberwithin{equation}{section}

\usepackage{amsmath,amssymb,enumerate}

\usepackage{epsfig,fancyhdr,color}

\usepackage{amssymb}
\usepackage{amsmath,amsthm}
\usepackage{bbm}
\usepackage{latexsym}
\usepackage{amscd}
\usepackage{psfrag}
\usepackage{graphicx}
\usepackage[all]{xy}

\usepackage[font=small,labelfont=bf]{caption}

\usepackage{mathtools}

\usepackage{tikz-cd}

\usepackage{stackrel}

\usepackage{mathabx,epsfig}

\usepackage[utf8]{inputenc}

\numberwithin{equation}{section}

\definecolor{NoteColor}{rgb}{1,0,0}

\definecolor{darkred}{RGB}{127,0,0}

\newtheorem*{theorem 1}{\rm\bf Proposition 1}
\newtheorem*{theorem 2}{\rm\bf Proposition 2}

\theoremstyle{definition}

\theoremstyle{remark}

\def\interieur#1{\mathord{\mathop{\kern 0pt #1}\limits^\circ}}

\begin{document}

\title[On the origins of Omicron]{On the origins of the Omicron variant of the SARS-CoV-2 virus}
\
\author{Robert Penner}
\address {\hskip -2.5ex Institut des Hautes \'Etudes Scientifiques\\
35 route des Chartres\\
Le Bois Marie\\
91440 Bures-sur-Yvette\\
France\\
{\rm and}~Mathematics Department,
UCLA\\
Los Angeles, CA 90095\\USA}
\email{rpenner{\char'100}ihes.fr}
\\
\author{Minus van Baalen}
\address {\hskip -2.5ex  CNRS \& Institut de Biologie de l'\'Ecole Normale Sup\'erieure\\
Paris Science Lettres\\
46 rue d'Ulm\\
75005 Paris\\
France\\}
\email{minus.van.baalen{\char'100}bio.ens.psl.eu}
\

\thanks{RP is the corresponding author, and it is a pleasure for him to thank Charles Swerdlow for critical comments.}

 \date{\today}


\begin{abstract}
A possible explanation based on first principles for the appearance of the Omicron variant of the SARS-CoV-2 virus is proposed
involving coinfection with HIV.  The gist is that the resultant HIV-induced immunocompromise
allows SARS-CoV-2 greater latitude to explore its own mutational space.  This latitude is not without
restriction, and a specific biophysical constraint is
explored.  Specifically, a nearly two - to fivefold discrepancy in backbone hydrogen bonding is observed
between sub-molecules in Protein Data Bank files of the spike glycoprotein yielding two conclusions: mutagenic residues in the 
receptor-binding subunit of the spike much more frequently do not participate in backbone hydrogen bonds;
and a technique of viral escape is therefore to remove such bonds within physico-chemical and functional constraints.
Earlier work, from which the previous discussion is entirely independent, explains these phenomena
from general principles of free energy, namely, the metastability of the glycoprotein.  The conclusions therefore likely hold more generally as principles in virology.
\end{abstract}

\maketitle



\section*{Introduction}

The Omicron variant of the SARS-CoV-2 virus has very recently arisen to global concern, owing to the
unprecedented number of mutations in its spike glycoprotein and their locations in key regions governing
both transmissibility and antibody binding.  In fact, there are 36 such mutations (see Table \ref{table:table})
according to the announcement of the proteome by Tulio de Oliveira, the director of the South African Center for Epidemic Response and Innovation.  This is in contrast to the 2 to 8 spike mutations typical for earlier Variants of Interest or Concern or Under Monitoring, for example with 5 such mutations for the Delta
variant \cite{outbreak}, 2 of which are in common with Omicron.   As is customary, these mutations are recorded relative to a specific initial variant called Wuhan-Hu-1 (UniProt Code P0DTC2).

One powerful aspect of mRNA or virus-vector vaccines is the ability to quickly develop multivalent vaccine cargos tailored to multiple viral variants using the same delivery system but with modified cargo, presumably with the same profile of side-effects.  Moderna, BioNTech and Johnson and Johnson have all announced just that, with vaccines targeting Omicron expected by early 2022, if necessary.  But it is facile to be confident that this will surely work, though
erna/NIAID vaccines, which deliver different mRNA instructions for exactly the same 2P stabilized spike, and yet present different efficacies with different side-effect profiles according to the CDC website, highlighting the subtlety of these issues.  
one can at least be hopeful that such modifications will stay abreast of the mutating viral genome.

\smallskip 

It is clearly important to understand the genesis of this latest mutational leap to Omicron, and this is the main topic of the current short note.

\smallskip

There are two key points we shall develop: one is a plausible scenario, which has its roots in the theory of speciation, involving coinfection by Covid-19 and HIV/AIDS leading to Omicron, as discussed in \cite{MsomiEtal:2021}; and another is the crucial energetic role that the Delta variant seems to have played in the evolution from Wuhan-Hu-1 to Delta to Omicron.  The first point is speculative, for there can be no certainty of the specific biological circumstance that led to Omicron though our argument is based upon well-accepted first principles.
The second point, however, is substantiated by explicit data involving an elementary counting of
hydrogen bonds in structures from the Protein Data Bank (PDB) \cite{PDB}; this second point
is furthermore explained by general considerations of free energy for metastable proteins.

\section{Results}

\subsection{Holey adaptive {landscapes} }

The highest national prevalence of about 20\% of HIV infection among the general population occurs in South Africa \cite{HIVSA}, (compared to less than 1\% globally \cite{HIVglobal}), with a corresponding SARS-CoV-2 infection rate of about 5\% (compared to about 3.5\% globally) \cite{outbreak} and presumed double-vaccination rate of under 25\% (compared to over 40\% globally).  It follows that there have been and are vast numbers of people in southern Africa coinfected with 
{HIV and SARS-CoV-2}.

Typically in viral coinfection \cite{virus} whether by various strains or species, the usual antigenic drift of a subject genome such as SARS-CoV-2 is accelerated via antigenic shift.
{We are aware of no evidence of this so far in the current case, but the presence of a coinfecting retrovirus such as HIV 
(with its machinery for address, transport, pore-forming, restriction and insertion of DNA that is 
not discriminating as to cargo) may enhance the scope for recombination. 
}

But it is a different interaction between HIV and SARS-CoV-2 that we highlight here.  Namely, a pernicious aspect of HIV/AIDS is the immediate burden placed on the immune system by the rapidly mutating HIV virions
and the concomitant overwhelmed host immune system that typically advances the mortal blow to AIDS patients {\cite{HIVvariability}}.  

In any case, the HIV-infected host immune system is weakened and becomes ineffective. This allows another coinfecting pathogen greater latitude to explore its own mutational landscape without being
terminated by the enfeebled immune system.  Surely there are still biophysical constraints on what
mutations are possible, and we shall confront one in the next subsection.  Such mutational latitude
can allow for large jumps in genome, as we have just witnessed in the genesis of
the Omicron variant.

An analogy from speciation is useful, for imagine a species that can feed and procreate with no particular
effort or skill.  In the course of generations with this lack of selective pressure, the genome is free
to explore countless mutations with no particular penalty or profit from any one over any other.
All bets are off, and huge leaps are possible through the mutational landscape over time.

So it is for SARS-CoV-2 in a host with weakened immune system.  Given sufficiently many generations
of viral replication over sufficiently many coinfected hosts, eventually there will be large deviation
from the original genome with {potential for} increased fitness.

This is a possible explanation for the monumental leap to the Omicron variant.
Such leaps have been discussed in the general context of speciation, for instance in \cite{holey}, from which we quote
``Rapid speciation is most likely for populations that are subdivided into a large number of small subpopulations." This is evidently the case in this viral analogy on several different scales.

{Thus, a host population with strong immune systems is likely to effectively corral the virus population in phenotype space where structural (as discussed below) and other constraints limit antigenic drift. However with sufficiently many hosts with weakened immune systems, the virus population can explore more and further in this phenotypic landscape, eventually leading to sudden antigenic leaps.}

This explanation seems more plausible than the classic antigenic shift in this particular case
since the mutations that define Omicron are essentially point mutations and insertions or
a couple of isolated deletions of a few contiguous residues.  In antigenic shift, one typically expects a series of 
backbone-contiguous recombinations, though the situation with a retrovirus may be more subtle as was already mentioned.

One consequence of our explanation for the genomic leap to Omicron is that this leap began from
Delta, which accounted for nearly all of the new infections in southern Africa before the advent of
Omicron.  The smattering of other circulating variants do not allow sufficient numerical momentum for the
mechanism we have just described.  Further evidence for this Delta to Omicron evolution and a discussion of Beta to Omicron will be
given in the next section.

\subsection{Wuhan-Hu-1 to Delta to Omicron}

Despite the latitude just discussed of viral mutation in a subject coinfected with HIV and SARS-CoV-2, there are also apparently certain fundamental constraints imposed by physics and chemistry.   At a basic minimum, there must not be so much free energy as to explode the molecule (typically for proteins on the order of 8-10 kcal/mole at laboratory conditions) nor so little as to compromise the metastability and hence function of the spike glycoprotein.  More stringent constraints on the considered mutation are imposed by requiring increased fitness of the variant in the host population, but this two-fold energetic constraint on
free energy is sacrosanct. 

Here we discuss one particular piece of empirical evidence of this energetic constraint, which was first described in \cite{pennerjcb1} for the Alpha-Gamma variants, then extended in \cite{pennervax} to include all Variants of Concern or Interest or under Monitoring in the PDB before the advent of Omicron.   Here we extend the discussion to include Delta and Omicron.

In order to state this finding, recall \cite{AFbook}  that
in any protein, hydrogen bonds form between backbone 
Nitrogen atoms N$_i$-H$_i$ and Oxygen atoms O$_j$=C$_j$ in different peptide units, and these
are called Backbone Hydrogen Bonds (or BHBs). 
(To be precise: A DSSP \cite{DSSP} hydrogen bond is accepted as a BHB provided that furthermore the distance between H$_i$ and O$_j$ is less than 2.7 \AA~ and $\angle NHO$ and $\angle COH$ each exceed 90$^\circ$.)
A protein residue R$_i$ itself is said to {\sl participate}
in a BHB if either the nearby Nitrogen N$_i$-H$_i$ donates to or the nearby Oxygen O$_i$=C$_i$ accepts
a BHB.  (Again to be precise, if at least two monomers of the trimeric spike participate, then the
residue itself participates.) In general in a protein, it is the BHBs that strongly contribute to the stability of 
secondary, tertiary and quaternary structures.  On average for all proteins, roughly 70-80\% of all residues participate in BHBs \cite{AFbook}.

Our novel finding on the mutagenesis of viral glycoproteins is:

\medskip

\leftskip .3in
\rightskip.3in

\noindent {\bf Basic Finding}: In the receptor-binding subunit of the spike glycoprotein, 
it is the residues which do not participate in BHBs that more frequently
mutate.

\leftskip=0ex\rightskip=0ex

\medskip

\leftskip .3in
\rightskip.3in

\leftskip=0ex\rightskip=0ex

\medskip

\noindent Our theory \cite{pennerjcb1,pennerjcb2,pennercmb,pennervax,naturepaper} of backbone free energy (BFE),
which is based in its essence on the geometry of the protein backbone,
assigns an absolute free energy in kcal/mole to each BHB, as specified by the 3d structure of the protein as determined by a PDB file.  The most trivial way
to avoid upsetting molecular BFE metastability is to mutate only those residues which do not participate in BHBs, since they are entirely neutral for BFE.  Insofar as this explanation for our Basic Finding is completely general, it seems reasonable to expect analogous higher mutagenic pressure on residues which are free from BHBs more generally for other metastable viral glycoproteins.

We shall quantify the Basic Finding for the three mutational steps Wuhan-Hu-1 to Delta,
Wuhan-Hu-1 to Omicron, and Delta to Omicron.  To this end, we 
introduce terminology from \cite{pennervax} as follows:  A residue is {\sl missing} if it is not modeled in the PDB file (which we argue can be interpreted as disorganized for high quality PDB files), it is {\sl absent} if it occurs in the PDB file but does not participate in either nearby backbone hydrogen bond (along the backbone), and it is {\sl unbonded} if it is neither missing nor absent in at least 5 of the PDB files in the database for that residue\footnote{Our findings are based upon the 15 high-quality PDB files {\tiny 6VXX 6X29 6X7 6XLU 6XM0 6XM3 6XM4 6ZB5 6ZGE 7A4N 7AD1 7DDD 7DF3 7DWY 7JWY} for Wuhan-Hu-1, and the 10 very good quality PDB files {\tiny 7V7O 7V7P$\ldots$7V7V} for the Delta variant. The PDB files {\tiny 7LYK 7LYL}$\ldots${\tiny 7LYQ} for the Beta variant B.1.351 are of a lower quality, so their interpretation is problematic
and not fully presented here; however the resulting $\bar{\rm S}_1$-entry for Beta to Omicron 
in Table \ref{table:table}  is 48\%.}.

Let ${\rm S}_1$ denote the sub-molecule of the spike S from N-terminus to the furin cleavage point at residue 681, and let ${\rm S}_2$ denote the complementary C-terminal molecule, respectively the receptor-binding and fusion subunits of S.  Several trends in any of the three mutational steps we study
are expected for reasons of location and function, namely:
${\rm S}_1$ is more disorganized than ${\rm S}_2$ (i.e.,  \#~missing is larger);
there are more loops in ${\rm S}_1$ than ${\rm S}_2$ (i.e., \# absent is larger);
and the average BFE of ${\rm S}_1$ is larger than ${\rm S}_2$.

Table \ref{table:table} quantifies our Basic Finding under various scenarios, with the
${\rm W}\mapsto(*)$ and $\Delta\mapsto O$ mutations comparable.  The ${\rm W}\mapsto O$
transition is anomalous with its much smaller percentage of unbonded mutated
residues.  The explanation follows from the last column, showing the large percentages of residues
that are bonded in W but not in $\Delta$, and hence of higher mutagenic potential for their transition
to $O$ according to the Basic Finding. 

\captionsetup{width=18cm}

\begin{table}[tbhp]
\centering
\title{\bf Percentage Unbonded Residues}
\begin{tabular}{cccccccccccccc}\\
$\underline{\mbox{\tiny \bf molecule}}$	&	$\underline{\mbox{\tiny\bf W$\mapsto$(*)}}$	& 	$\underline{\mbox{\tiny\bf $\Delta\mapsto O$}}$	&	$\underline{\mbox{\tiny\bf W$\mapsto O$}}$	&
$\underline{\mbox{\tiny\bf W$\mapsto\Delta$}}$ &	&				\\\\

S&35&40&35&30\\
$\bar{\rm S}$&61&61&22&42\\\\

S$_1$&42&46&42&32\\
$\bar{\rm S}_1$&74&67&27&43\\\\

S$_2$&25&32&25&28\\
$\bar{\rm S}_2$&15&33&0&33\\

\hline
\end{tabular}
\caption{\tiny All table entries are percentages of unbonded residues in each molecule. 
W, $\Delta$, and $O$ are respectively the Wuhan-Hu-1, Delta and Omicron variants,
and (*) denotes the collection of Variants of Concern or Interest or under Monitoring \cite{outbreak}
before the advent of Omicron with combined mutated residues
(5 9 12 18-20 26) 52 67 69-70 75-76 80 95 136 138 144-145
152 156-158 190 215 243-244 246-253 346 417 449 452
478 484 490 501 570 614 641 655 677 679 681 701 716
796 859 888 899 950 982 1027 1071 1092 1101 1118 (1176), where the residues in parentheses are
outside PDB-coverage and are not reflected in this table.  It is Tulio de Oliveira's preliminary collection
of mutated residues 67 69 70 142-145 211 212 214 339 371 373 375 417 440 446 477 478 484 493 496 498 501 505 547 614 655 679 681 764 796 856 954 969 981 that serve to define $O$ for our purposes.  We let
$\bar{\rm S}={\rm S}\cap M$, $\bar{\rm S}_1={\rm S}_1\cap M$, $\bar{\rm S}_2={\rm S}_2\cap M$ respectively denote the collection of mutagenic residues $M$ in each of the molecules S,${\rm S}_1$,${\rm S}_2$, with $M$ determined by (*) in the second column and by $O$ in the third and fourth columns.  The last column is simply the percentage of residues which are bonded in W and unbonded in $\Delta$ in each molecule.
}
\label{table:table}
\end{table}

\section{Discussion}

We have proposed a plausible 
{explanation} for the genomic leap from the Delta to the Omicron variants
of SARS-CoV-2 based upon coinfection with HIV.  In effect, HIV weakens the host immune system,
which allows SARS-CoV-2 to more freely explore its mutational landscape, as in certain models of speciation.
With sufficiently many replications within the large coinfected population in southern Africa, a mutation 
{that allows the virus to escape from the corraling effect of its host immune system}
would be bound to arise eventually.  

However, this greater mutational latitude is still not without biophysical constraints,
one such being our Basic Finding that residues in the receptor-binding
subunit ${\rm S}_1$ of the spike have higher mutagenic pressure if they do not participate in backbone hydrogen bonds.  This is in keeping with general principles of free energy and follows from metastability of the spike.

This finding does not hold for the fusion subunit ${\rm S}_2$, possibly because the underlying PDB files reflect the prefusion conformation of the spike at relatively higher pH than the fusion subunit activation and because these PDB files model the spike molecule pre-cleavage with the associated steric constraints imposed by ${\rm S}_1$ sitting as a cap on top of ${\rm S}_2$.

Assuming the Basic Finding, the transition from the original Wuhan-Hu-1 variant directly to the
Omicron variant seems most unlikely, with the Delta variant a seemingly necessary intermediary.
The dynamics of this is that the transition from Wuhan-Hu-1 to Delta removes a multitude of backbone hydrogen bonds thereby allowing residues to further mutate to Omicron in keeping with the Basic Finding.

\section{Conclusion} 

Our findings admit explanation by general principles, and so may hold more generally, to wit:
weakened immune systems allow pathogens to capitalize on large mutational leaps;
being free from backbone hydrogen bonds increases the mutagenic potential within the receptor-binding subunit of a viral glycoprotein; deleting backbone hydrogen bonds, within the constraints of molecular
functionality, can increase the mutagenic potential of a glycoprotein.




\begin{thebibliography}{ABCD}

\bibitem{PDB}  Berman, H.M., Westbrook, J., Feng, Z., et al. 2000. The Protein Data Bank. {\it Nucleic Acids Research} 28, 235--242. Available online: {\tt http://www.rcsb.org/pdb/}.

\bibitem{virus} Dimmock, N.J. et al.~(2007). Introduction to Modern Virology, 6th edition. Blackwell, Oxford, UK.

\bibitem{HIVSA}
Dwyer-Lindgren, L. et al.~(2019). {\it Nature} {\bf 570}, 189--193.




\bibitem{AFbook}
Finkelstein, A.V., and Ptitsyn, O. 2016. Protein Physics, A Course of Lectures. 2nd edition. Academic Press, London, UK.

\bibitem{holey}
Gavrilets, S., 1999
A Dynamical Theory of Speciation on Holey
Adaptive Landscapes, {\it American Naturalist} {\bf 154}, 1--22.


\bibitem{DSSP} 
Kabsch, W. and Sander, C. 1983, 
Dictionary of protein secondary structure: Pattern recognition of hydrogen-bonded and geometrical features, 
{\it Biopolymers} {\bf 22}, 2577--637. Available online: {\tt https://swift.cmbi.umcn.nl/gv/dssp/}.

\bibitem{MsomiEtal:2021}
Msomi, N., Lessells, R., Mlisana, K., and de~Oliveira, T. (2021).
\newblock {Africa: tackle HIV and COVID-19 together}.
\newblock {\em Nature}, 600:33--66.

\bibitem{outbreak}
Mullen, J.L., Center for Viral Systems Biology, et al. 2020,
outbreak.info. Available online: {\tt https://outbreak.info/}.

\bibitem{HIVvariability}
Nowak, M.~A. (1992).
\newblock Variability of {HIV} infections.
\newblock {\em J. Theor. Biol.}, 155:1--20.



\bibitem{pennerjcb1}{
{Penner, R.} 2020
{Backbone Free Energy Estimator Applied to Viral Glycoproteins},
{\it Journal of Computational Biology},
{\bf 27},
{10},
{1495--1508},
{Liebert}.
}

\bibitem{pennerjcb2}{
{Penner, R.} 2020
 {Conserved High Free Energy Sites
in Human Coronavirus Spike Glycoprotein Backbones},
{\it Journal of Computational Biology},
{\bf 27},
{11},
{1622--1630},
{Liebert}.
}

\bibitem{pennercmb}
{Penner, R.} 2021,
Antiviral Resistance against Viral Mutation: Praxis and Policy for SARS-CoV-2,
{\it Computational and Mathematical Biophysics} {\bf 9}(1), 81--89.

\bibitem{pennervax}
{Penner, R.} 2021,
Mutagenic distinction between the receptor-binding and fusion subunits of the SARS-CoV-2 spike glycoprotein,
bioRxiv 2021.11.15.468283; doi: https://doi.org/10.1101/2021.11.15.468283.




\bibitem{naturepaper}{
{Penner, R., et al.} 2014,
{Hydrogen bond rotations as a uniform structural tool for analyzing protein architecture},
{\it Nature Communications},
{\bf 5},
{5803},
{Springer-Nature}.  Available online: {\tt https://bion-server.au.dk/hbonds/}
}

\bibitem{HIVglobal}
UNAIDS. 2021. The Global HIV/AIDS Epidemic.
{\tt https://www.unaids.org/en/resources/fact-sheet}.

 
 
 \end{thebibliography}
\end{document}